\documentclass[12pt]{iopart}
\usepackage{iopams}
\usepackage{graphicx}
\usepackage{color}

\begin{document}

\newcommand \be {\begin{equation}}
\newcommand \ee {\end{equation}}
\newcommand \bea {\begin{eqnarray}}
\newcommand \eea {\end{eqnarray}}
\newcommand \la {\langle}
\newcommand \ra {\rangle}
\newcommand \ve {\varepsilon}
\newcommand{\mC}{\mathcal{C}}
\newcommand{\mT}{\mathcal{T}}
\newcommand{\mE}{\mathcal{E}}
\newcommand{\mP}{\mathcal{P}}
\newcommand{\mR}{\mathcal{R}}
\newcommand{\mA}{\mathcal{A}}
\newcommand{\Lf}{\mathcal{L}}
\newcommand{\p}[1]{\left(#1\right)}

\title[]{Stochastic resetting of a population of random walks with resetting-rate-dependent diffusivity}

\author{Eric Bertin}

\address{LIPHY, Univ.~Grenoble Alpes and CNRS, F-38000 Grenoble, France}
%\ead{eric.bertin@univ-grenoble-alpes.fr}

\begin{abstract}
  We consider the problem of diffusion with stochastic resetting in a population of random walks where the diffusion coefficient is not constant, but behaves as a power-law of the average resetting rate of the population. Resetting occurs only beyond a threshold distance from the origin. This problem is motivated by physical realizations like soft matter under shear, where diffusion of a walk is induced by resetting events of other walks.
  We first reformulate in the broader context of diffusion with stochastic resetting the so-called H\'ebraud-Lequeux model for plasticity in dense soft matter, in which diffusivity is proportional to the average resetting rate.
Depending on parameter values, the response to a weak external field may be either linear or non-linear with a non-zero average position for a vanishing applied field, and the transition between these two regimes may be interpreted as a continuous phase transition.
 Extending the model by considering a general power-law relation between diffusivity and average resetting rate, we notably find a discontinuous phase transition between a finite diffusivity and a vanishing diffusivity in the small field limit.  
\end{abstract}

\section{Introduction}

Many stochastic processes consist of a combination of a continuous diffusive dynamics and discontinuous stochastic jumps. This is the case in particular for search processes which typically become more efficient by making random jumps to explore distant areas in a shorter time \cite{Benichou11}.
A simple and paradigmatic model for such intermittent diffusive dynamics is the diffusion process with stochastic resetting
\cite{Evans11PRL,Evans11JPA,Montero13,Montero16} ---see \cite{Evans20rev} for a review.
At odds with standard random walks, a random walk with stochastic resetting to the origin converges to a stationary statistical state even in an unbounded domain \cite{Mendez16,Eule16}.
This minimal model has been extended in many different directions, including 
arbitrary spatial dimensions \cite{Evans14}, bounded domains \cite{Christou15},
Langevin dynamics \cite{Gupta19}, space-dependent diffusivity \cite{Sandev22}, time-dependent resetting rate \cite{Pal16} or non-Poissonian resetting dynamics \cite{Nagar16}.
Anomalous diffusion with stochastic resetting has also been considered
\cite{Masoliver19,Stanislavsky22}, notably in the context of record statistics \cite{Majumdar22}.
The diffusion process with stochastic resetting also fostered further works on search strategies \cite{Chechkin18,Schumm21,Mercado21} and their optimality \cite{Kusmierz14}.

Beyond these different generalizations, another natural extension is to consider interacting random walks with resetting, as done recently for instance in the context of population genetics \cite{Silva22}.
Although interactions may occur in many different ways, like for instance non-crossing conditions, a physically-motivated type of interaction is to consider that in practical realizations, resetting events often lead to energy dissipation and may thereby generate some noise. A large population of random walks with resetting may thus continuously generate noise through resetting events, and this noise may be (fully or partly) the physical source of diffusion of the walks. This mechanism is at play in the so-called elastoplastic scenario for the deformation of soft amorphous materials \cite{Nicolas18RMP}. In such systems, the local mechanical stress performs a random walk with stochastic resetting (resetting corresponds here to a local stress relaxation called plastic event) once a stress threshold is overcome. Since such systems are athermal, the only source of stress diffusion is the mechanical noise generated by distant plastic events and transmitted through a long-range elastic propagator \cite{Nicolas18RMP}.
A mean-field model of this elastoplastic scenario, called the H\'ebraud-Lequeux (HL) model, has been proposed more than twenty years ago \cite{HL98}. It basically consists of a population of random walks with stochastic resetting beyond a threshold distance to the origin, and such that the diffusion coefficient is proportional to the average resetting rate of the population. This dependence of the diffusion coefficient on the average resetting rate results in effective interactions between the walks.

The goal of this paper is twofold. First, we aim at reformulating in the broader context of diffusion with stochastic resetting the results of the HL model known in the specific context of the deformation of soft amorphous materials. 
Second, we generalize the results of the HL model by considering a more general relation between diffusivity and average resetting rate. This generalization leads to a rich phenomenology that we discuss here.
We believe that such a physically-motivated way to introduce mean-field interactions between random walks with stochastic resetting could be of interest to the community working on this topic, and might lead to a number of further developments in the field.

The paper is organised as follows. Sec.~\ref{sec:description} introduces the HL model and briefly discusses its interpretation in a soft matter context.
Then Sec.~\ref{sec:evol:proba} uses the derivation of the stationary probability distribution to obtain a self-consistent equation satisfied by the field-dependent diffusion coefficient. Finally Sec.~\ref{sec:response} evaluates the diffusion coefficient in the small field limit as a function of model parameters, leading to the identification of distinct regimes, either linear or non-linear, for the average position of the walk as a function of the external field. Sec.~\ref{sec:conclusion} eventually draws some conclusions.

\section{Description of the model}
\label{sec:description}

\subsection{A population of random walks with stochastic resetting}

We consider a large population of $N$ random walks with stochastic resetting in one dimension. Each walk is described by its position $x_i$ ($i=1,\dots,N$), obeying the Langevin equation
\be \label{eq:Langevin}
\frac{dx_i}{dt} = h + \xi_i(t)
\ee
where $h$ is the applied external field and $\xi_i(t)$ is a Gaussian white noise satisfying
\be
\la \xi_i(t) \xi_j(t')\ra = 2D(t) \, \delta_{ij}\, \delta(t-t'),
\ee
with a time-dependent diffusion coefficient $D(t)$.
The continuous Langevin evolution described by Eq.~(\ref{eq:Langevin}) is supplemented by a random resetting rule corresponding to a stochastic jump to $x_i=0$, with a position-dependent transition rate $\lambda(x_i)$.
In the following, we restrict ourselves to the functional form
\be \label{eq:def:lambdax}
\lambda(x) = \lambda_0 \, \theta(|x|-b),
\ee
with $\lambda_0>0$ a constant rate, $b>0$ a threshold distance, and $\theta(x)$ the Heaviside function, equal to $\theta(x)=1$ for $x\ge 0$ and $\theta(x)=0$ for $x< 0$. In the following, we set $\lambda_0=1$ and $b=1$ by choosing appropriate time and length units.

Up to now, the $N$ random walks are statistically independent. The idea is to introduce a mean-field coupling between them by choosing the diffusivity $D(t)$
to be a function of the average resetting rate $\Gamma(t)$ defined as
\be \label{eq:def:Gamma}
\Gamma(t) = \int_{-\infty}^{\infty} \lambda(x)\, p(x,t)\, dx,
\ee
where $p(x,t)$ is the probability distribution of the position $x$ of a random walk at time $t$. In the infinite $N$ limit, $\Gamma(t)$ precisely corresponds to the resetting rate of the population of walkers.
In the following, we consider for definiteness the functional dependence
\be \label{eq:rel:D:Gamma}
D(t) = \alpha \, \Gamma(t)^{\beta},
\ee
with $\alpha>0$ and $\beta>0$. The limiting case $\beta=0$ corresponds to the usual diffusion with stochastic resetting problem, where the $N$ walkers are statistically independent. We exclude the case $\beta<0$ because we aim at describing a physical situation where the diffusivity of a given walker is induced by resettings of other walks in the population, and thus one should have $D(t)\to 0$ when $\Gamma(t) \to 0$.

\subsection{The HL model}

As mentioned in the introduction, a physically grounded implementation of the above model corresponds to the HL model \cite{HL98,EPJE15,Nicolas18RMP,Bertin16rev,Bouchaud16}, which describes in a mean-field way the plastic deformation of dense soft amorphous materials \cite{Puosi15}
(see also \cite{Gati06a,Gati06b,Olivier10,Olivier11,Olivier12} for more mathematically oriented studies of the HL model, and \cite{Olivier13} for an extension to higher dimensions). In such materials, deformation occurs via localised plastic events which release stress \cite{Nicolas18RMP}. Such events occur when the local stress exceeds a threshold, in analogy to the transition rate $\lambda(x)$ defined in Eq.~(\ref{eq:def:lambdax}).
Quite importantly, in finite-dimensional systems the locally released stress is redistributed throughout the system by a long-range elastic propagator \cite{Nicolas18RMP}.
A peculiarity of this propagator is that it is anisotropic and takes either positive or negative values depending on the direction considered.
In an elementary mean-field scenario, one may divide the system into boxes with a volume comparable to the volume of a rearranging region, and treat the effect of the stress redistribution process as a Gaussian white noise.
This is the assumption made in the HL model \cite{HL98}, which consistently assumes $\beta=1$ in relation (\ref{eq:rel:D:Gamma}), that is a proportionality between the diffusivity $D$ and the average resetting rate $\Gamma$.
Eq.~(\ref{eq:rel:D:Gamma}) may also be generalized by adding a constant term on the right hand side. Note that this term would model an additional source of noise (e.g., active noise \cite{Matoz17}) that does not depend on the average resetting rate.

%Before going further into technical aspects, note that detailed studies
%of the HL model \cite{Gati06a,Gati06b,Olivier10,Olivier11,Olivier12} and of its generalization to higher space dimensions \cite{Olivier13} can also be found in the mathematical literature.

\section{Evolution of the probability distribution}
\label{sec:evol:proba}

In the limit $N\to\infty$, the population of random walks can be described by a non-linear evolution equation for the probability distribution $p(x,t)$,
\be \label{eq:evol:pxt}
\frac{\partial p}{\partial t} = -h \frac{\partial p}{\partial x}
+ D(t) \frac{\partial^2 p}{\partial x^2} - \lambda(x)\, p(x,t) + \Gamma(t) \delta(x)
\ee
with $\delta(x)$ the Dirac delta distribution, and where $\lambda(x)$, $\Gamma(t)$ and $D(t)$ are defined in Eqs.~(\ref{eq:def:lambdax}), (\ref{eq:def:Gamma}) and (\ref{eq:rel:D:Gamma}) respectively.
The stationary solution $p_{\rm st}(x)$ is determined by first considering $D$ and $\Gamma$ as given constants, and then determining them in a self-consistent
way at the end of the calculation.
On each of the intervals $(-\infty,-1)$, $(-1,0)$, $(0,1)$ and $(1,+\infty)$,
Eq.~(\ref{eq:evol:pxt}) reduces to a linear second order ordinary differential equation for $p_{\rm st}(x)$. The different integration constants appearing after integration on each interval are fixed by (i) using appropriate boundary conditions between intervals at $x=-1$, $0$ and $1$, (ii) by ensuring that $p_{\rm st}(x)$ goes to zero when $x\to \pm\infty$, and (iii) by using the normalization condition $\int_{-\infty}^{\infty} p_{\rm st}(x)\, dx=1$.
The boundary conditions in $x=-1$ and $1$ are that the probability $p_{\rm st}(x)$ is continuous, and that the probability flux
\be \label{eq:def:Jx}
J(x) = h p_{\rm st}(x) - D \frac{dp_{\rm st}}{dx}
\ee
is continuous (because of its diffusive nature), which implies the continuity of the derivative $dp_{\rm st}/dx$.
At $x=0$, the probability distribution is also continuous, but not its derivative, because of the resetting probability flux $\Gamma$. One can write the corresponding probability flux balance at $x=0$,
\be
J(0^+) - J(0^-) = \Gamma,
\ee
which using Eq.~(\ref{eq:def:Jx}) yields an explicit condition on the discontinuity of the derivative $dp_{\rm st}/dx$ at $x=0$ (more details can be found in \cite{EPJE15}).
Having determined the stationary distribution $p_{\rm st}(x)$ for fixed values of the diffusivity $D$ and the external field $h$, one can express the average resetting rate $\Gamma$ as a function of $D$ and $h$ as
\be
\Gamma(D,h) = \frac{D}{f(D,h)},
\ee
where the function $f(D,h)$ is given by \cite{EPJE15}
\be
f(D,h) = D \left[1+ \frac{1+\left(\sqrt{1+\frac{4D}{h^2}}+\frac{2D}{|h|}\right) \, \tanh \frac{|h|}{2D}}{|h|\left(\tanh \frac{|h|}{2D} + \sqrt{1+\frac{4D}{h^2}}\right)}\right] \,.
\ee
The value of $D$ is then determined self-consistently using Eq.~(\ref{eq:rel:D:Gamma}), leading to a closed equation on $D$:
\be \label{eq:self:consist}
f(D,h) = \alpha^{1/\beta} D^{1-1/\beta}.
\ee
In principle, Eq.~(\ref{eq:self:consist}) should be solved by determining its solution $D(h)$ for any fixed external field $h$. In practice, such a resolution for an arbitrary value of $h$ can only be performed numerically.
However, as discussed below, it is possible to determine analytically $D(h)$ to leading order in the limit $h\to 0$.

Once $D(h)$ is determined, the distribution $p_{\rm st}(x)$ is known, and one can evaluate arbitrary average observables. Our interest here goes more specifically to the average position $\la x\ra$ of the walker.
In the absence of external field, $h=0$, the average position $\la x\ra=0$ in the stationary state, because of the symmetry $x\to-x$.
For $h\ne 0$, $\la x\ra$ can be evaluated from the knowledge of $p_{\rm st}(x)$, and one finds after some algebra
\be \label{eq:response:Dcst}
\la x\ra = \frac{h}{12D}\; \chi(D)+O(h^3)\,,
\ee
where the function $\chi(D)$ reads as
\be
\chi(D) = \frac{1+4D^{1/2}+12D+24D^{3/2}+24D^2}{1+2D^{1/2}+2D}\,.
\ee
Since $\chi(D)\to 1$ when $D\to 0$, the average position $\la x\ra$ is proportional to $h/D$ when $h/D\ll 1$.

Now considering the fact that $D$ depends on $h$ when taking into account Eq.~(\ref{eq:rel:D:Gamma}), the average position $\la x\ra \sim h/D(h)$ may behave linearly or non-linearly with $h$ in the limit $h\to 0$, depending on whether $D(h)$ goes to a finite value or to zero in this limit ---provided $D(h)$ goes to zero slower than $h$ to fulfill the assumption $h/D\ll 1$ used to derive Eq.~(\ref{eq:response:Dcst}). We will see below that both situations may occur in the model depending on parameter values.

\section{Linear and non-linear response regimes}
\label{sec:response}

We focus in this section on the determination of the field-dependent diffusion coefficient $D(h)$ in the limit $h\to 0$ as a function of the two parameters $\alpha$ and $\beta$ introduced in Eq.~(\ref{eq:rel:D:Gamma}).
For the sake of clarity, we analyse separately the cases $\beta=1$, $\beta<1$ and $\beta>1$.

\subsection{Case $\beta=1$}

The case $\beta=1$ corresponds to the standard HL model, and we reformulate here in a more pedagogical way some of the results reported in \cite{HL98,EPJE15}, rephrasing them in the generic framework of random walks with stochastic resetting. In the case $\beta=1$, Eq.~(\ref{eq:self:consist}) allowing for the determination of $D$ simplifies to
\be \label{eq:self:consist:b1}
f(D,h) = \alpha\,.
\ee
To determine $D$ in the limit $h\to 0$, one may first take the limit $h\to 0$ in the function $f(D,h)$, and one finds:
\be \label{eq:self:consist:b1:h0}
f_0(D) \equiv \lim_{h\to 0} f(D,h) = \frac{1}{2}+\sqrt{D}+D\,.
\ee
Hence for any fixed $D>0$, $f_0(D) >\frac{1}{2}$. For $\alpha>\frac{1}{2}$, the equation $f_0(D)=\alpha$ thus admits a solution $D_0>0$ of Eq.~(\ref{eq:self:consist:b1}) in the limit $h\to 0$.
One can get the leading correction in $h$ of $D(h)$ by evaluating the first correction in $h$ to $f(D,h)$ when $h\to 0$, yielding
\be
f(D,h) = f_0(D) + h^2 f_1(D) + O(h^4)\,,
\ee
where the function $f_1(D)$ is given by
\be
f_1(D) = -\frac{1}{2} \left[\frac{1}{4\sqrt{D}}+\frac{1}{2D}
    +\frac{1}{3D\sqrt{D}} + \frac{1}{12D^2} \right].
\ee
Note that $f_1(D)<0$ for all $D>0$.
Expanding $D(h)$ as
\be \label{eq:regular:ansatz}
D(h)=D_0 + h^2 D_1 + O(h^4)
\ee
for $h\to 0$, one finds $D_1=-f_1(D_0)/f_0'(D_0)>0$.

Using Eq.~(\ref{eq:response:Dcst}), the average position is then given by
\be \label{eq:xav:regular}
\la x \ra = \frac{h}{12D_0}\; \chi(D_0) + O(h^3)\,.
\ee
The response to the external field is thus linear to leading order, with a regular (i.e., cubic) subleading correction that we do not evaluate explicitly ---this would require to compute the cubic response in Eq.~(\ref{eq:response:Dcst}). Note that the $h$-dependence of $D(h)$ does not modify here the linear response with respect to the case of a constant diffusion coefficient $D_0$; it only contributes to non-linear corrections at order $h^3$ and higher.

In contrast, when $0<\alpha<\frac{1}{2}$, the equation $f_0(D) = \alpha$ has no solution.
In this case, one has to come back to Eq.~(\ref{eq:self:consist:b1}) for finite $h$, and to look for a parametrization of $D(h)$ that goes to zero when $h\to 0$. Indeed, although $f_0(0)=\frac{1}{2}$ from Eq.~(\ref{eq:self:consist:b1:h0}), one has $\lim_{D\to 0}f(D,h)=0$ for all $h>0$, meaning that the limits $D\to 0$ and $h\to 0$ do not commute.
It follows that $f(D,h)$ actually reaches values lower than $\frac{1}{2}$ if one parameterises $D$ as a function of $h$ when taking the limit $h\to 0$. Since $D(h)$ is expected to go to zero when $h\to 0$, a natural parameterisation is to assume that $D(h)=u|h|^{\nu}$, with $\nu>0$ a given exponent, and $u$ the rescaled diffusion coefficient.
As a first trial, we investigate the case $\nu=1$.
Assuming $D=u|h|$, the function $f(D,h)$ can be expanded for $h\to 0$ as
\be
f(u|h|,h) = \phi_0(u) + \sqrt{|h|} \, \phi_1(u) + O(|h|)
\ee
with
\be
\phi_0(u) = u\tanh\frac{1}{2u}\,, \qquad
\phi_1(u) = \frac{\sqrt{u}}{2} \left[1+2u\tanh\frac{1}{2u}
  - \tanh^2\frac{1}{2u}\right].
\ee
Interestingly, the function $\phi_0(u)$ takes values in the range $(0,\frac{1}{2})$. Hence for $0<\alpha<\frac{1}{2}$, the equation $\phi_0(u)=\alpha$ has a solution $u_0>0$. Expanding $u(h)$ as $u(h)=u_0+\sqrt{|h|}\, u_1$, one finds
$u_1=-\phi_1(u_0)/\phi_0'(u_0)<0$.
Taking into account the relation $D=u|h|$ as well as the expression (\ref{eq:response:Dcst}) of $\la x \ra$, the average position is obtained to leading order in a $\sqrt{|h|}$ expansion as
\be \label{eq:HB}
\la x \ra = \left[\frac{1}{12u_0} + \frac{1}{6}\left( \frac{1}{\sqrt{u_0}} - \frac{u_1}{2u_0^2}\right) \sqrt{|h|}\right]\,
    {\rm sgn}(h) + O(h)\,.
    \ee
Hence for $0<\alpha<\frac{1}{2}$, the average position of the walk does not vanish in the limit $h\to 0$. This comes from the fact that the diffusion coefficient also goes to zero as $D \propto |h|$, which effectively enhances the bias generated by the external field $h$.
The subleading correction, proportional to $\sqrt{|h|}$ is also of interest because of its singular behaviour.
In the HL model for sheared soft amorphous materials, the average position is interpreted as the average mechanical stress in the material in response to a deformation rate given by $h$, and the behaviour given by Eq.~(\ref{eq:HB}) with the $\sqrt{|h|}$ correction is called the Hershel-Bulkley law \cite{EPJE15,Nicolas18RMP}.

Up to now, we have been able to solve Eq.~(\ref{eq:self:consist:b1}) in the small $h$ limit for the two cases $\alpha>\frac{1}{2}$ and $0<\alpha<\frac{1}{2}$ by using the scaling relations $D \propto |h|^{\nu}$ with $\nu=0$ and $\nu=1$ respectively. By doing so, the value $\alpha=\frac{1}{2}$ has been left aside.
It is thus natural to expect that Eq.~(\ref{eq:self:consist:b1}) may be solved with the ansatz $D \propto |h|^{\nu}$ for some intermediate value of $\nu$.
We thus set $D=u |h|^{\nu}$, with $0<\nu<1$ an exponent to be determined. A lowest order expansion of $f(u|h|^{\nu},h)$ for $h\to 0$ leads to
\be
f(u|h|^{\nu},h) = \frac{1}{2} + \sqrt{u} \, |h|^{\nu/2}
- \frac{h^{2-2\nu}}{24u^2} + o(|h|^{\gamma})\,,
\ee
with $\gamma = \max(\frac{\nu}{2},2-2\nu)$.
Assuming first that the two exponents $\frac{\nu}{2}$ and $2-2\nu$ differ, one finds no solution for Eq.~(\ref{eq:self:consist:b1}) with $\alpha=\frac{1}{2}$.
One then concludes that the two exponents must be equal, leading to $\nu=\frac{4}{5}$. It follows that $u=(24)^{-2/5}$ is solution of Eq.~(\ref{eq:self:consist:b1}) with $\alpha=\frac{1}{2}$ and $D=u |h|^{4/5}$.
The resulting average position $\la x \ra$ is then given for $h\to 0$ by
\be
\la x \ra = (432)^{-1/5} |h|^{1/5} {\rm sgn}(h)+o(|h|^{1/5}),
\ee
corresponding to a strongly non-linear response with a non-trivial exponent $\frac{1}{5}$.

In other words, one has for $\beta=1$ the equivalent of a phase transition as a function of the parameter $\alpha$, with a critical value $\alpha_{\rm c}=\frac{1}{2}$. The order parameter of the transition is the average position $\la x\ra$ of the walk. For $\alpha > \alpha_{\rm c}$, $\la x\ra \to 0$ when $h\to 0$ and there is no spontaneous symmetry breaking.
In contrast, for $\alpha < \alpha_{\rm c}$, $\la x\ra$ goes to a finite value
when $h\to 0$ [see Eq.~(\ref{eq:HB})], corresponding to a spontaneous symmetry breaking.
The order parameter vanishes at the critical point as
$\la x\ra \sim (\alpha_{\rm c}-\alpha)^{1/2}$ \cite{EPJE15}.
In this respect, the situation is similar to the mean-field Ising model, where the average magnetisation goes when $h\to 0$ to a non-zero value
%(whose sign is given by the sign of the magnetic field $h$ when $h\to 0$)
which behaves as a square-root of the distance to the critical point.
However, right at the critical point $\alpha = \alpha_{\rm c}$, the average position behaves as a power law of $h$, $\la x \ra \sim |h|^{1/5}\,{\rm sgn}(h)$, corresponding to a critical exponent $\delta=5$ in the usual notations of critical phenomena.
Interestingly, and although the present model is purely of mean-field type, the value $\delta=5$ differs from the standard exponent $\delta=3$ found in the mean-field Ising model.

\subsection{Case $\beta<1$}

\begin{figure}
  \begin{center}
    \includegraphics[width=0.6\linewidth]{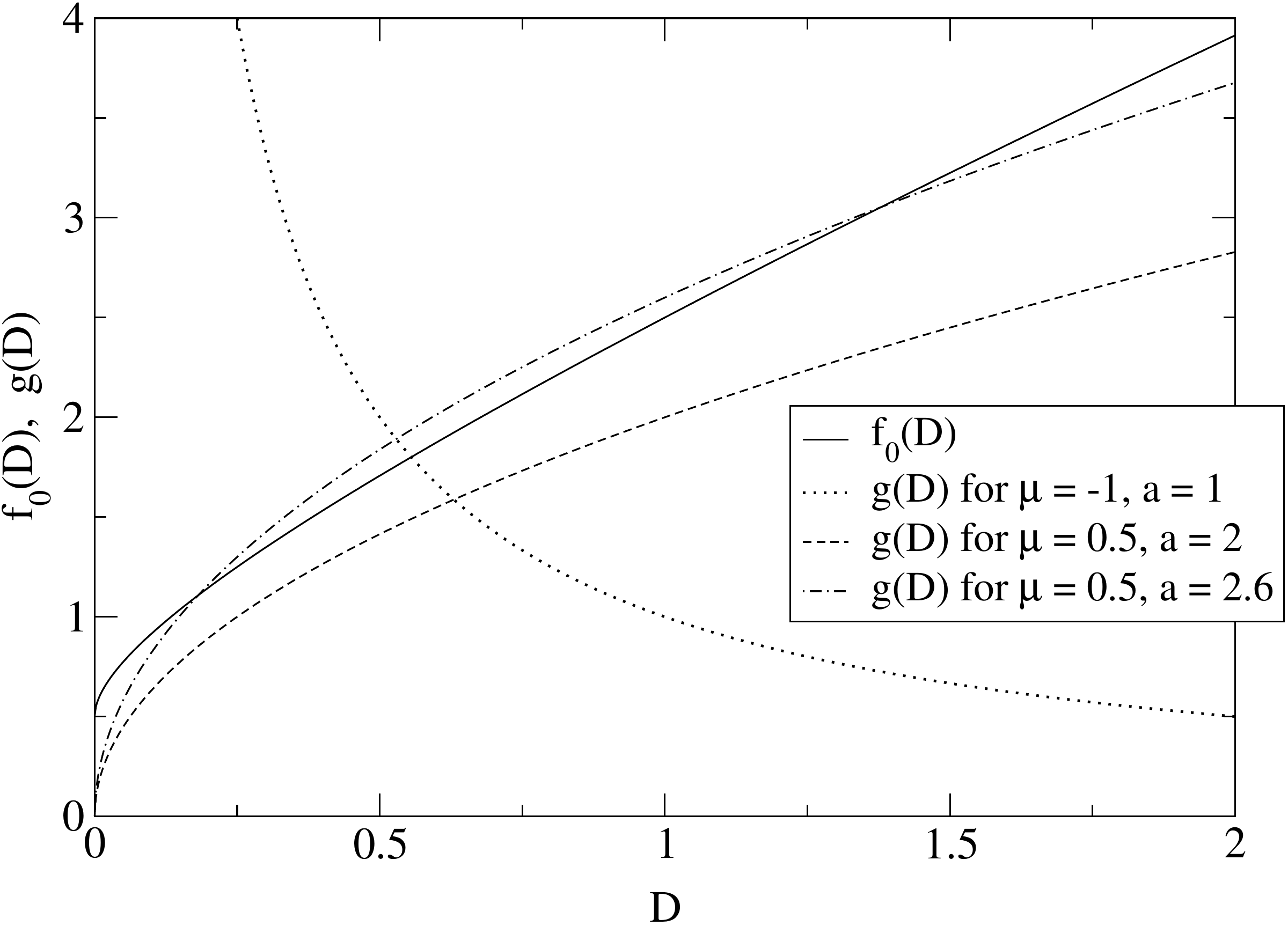}
    \caption{Graphical illustration of the solutions of the equation $f_0(D)=aD^{\mu}$, with $\mu=1-\frac{1}{\beta}$, obtained as the intersections of the curves representing the functions $f_0(D)$ (full line) and $g(D)=aD^{\mu}$ (other lines). For $\mu<0$ (i.e., $\beta<1$), the equation always has a single solution, illustrated here for $\mu=-1$ and $a=1$ ($g(D)$ is plotted as a dotted line). For $0<\mu<1$ (i.e., $\beta>1$), there exists a critical amplitude $a_{\rm c}$ such that the equations admits no solution for $a<a_{\rm c}$ and two solutions for $a>a_{\rm c}$. Illustration: $\mu=\frac{1}{2}$ with $a=2$ ($g(D)$ plotted as a dashed line) and $a=2.6$ (dot-dashed). The critical amplitude for $\mu=\frac{1}{2}$ is $a_{\rm c} \approx 2.414$.}
    \label{fig:scplot}
  \end{center}
\end{figure}

When $\beta \ne 1$, Eq.~(\ref{eq:self:consist:b1}) is replaced by
Eq.~(\ref{eq:self:consist}), which we rewrite for convenience as
\be \label{eq:self:consist:bne1}
f(D,h) = a \, D^{\mu}\,,
\ee
with
\be
a = \alpha^{1/\beta}, \qquad \mu = 1-\frac{1}{\beta}\,.
\ee
When $\beta<1$, one has $\mu<0$ and Eq.~(\ref{eq:self:consist:bne1}) boils down for $h\to 0$ to
\be \label{eq:D0:muneg}
\frac{1}{2}+\sqrt{D}+D = a \, D^{\mu}\,,
\ee
which always has a solution $D_0$ (see Fig.~\ref{fig:scplot}), that can be determined numerically.
Using the same expansion Eq.~(\ref{eq:regular:ansatz}) as in the case $\beta=1$, one finds for $D_1$,
\be \label{eq:D1:muneg}
D_1 = \frac{f_1(D_0)}{\mu a D_0^{\mu-1}-f_0'(D_0)}\,.
\ee
One then eventually obtains the same formal regular expansion in $h$ for the
average position $\la x\ra$ as in Eq.~(\ref{eq:xav:regular}),
but now with $D_0$ solution of Eq.~(\ref{eq:D0:muneg}).
Note that at odds with the case $\beta=1$, no transition occurs as a function of $\alpha$ in the case $\beta<1$.

\subsection{Case $\beta>1$}

We now have to solve Eq.~(\ref{eq:self:consist:bne1}) in the case $\beta>1$, which corresponds to $0<\mu<1$.
As illustrated on Fig.~\ref{fig:scplot}, this equation may either have zero solution below a critical amplitude, $a<a_{\rm c}$, or two solutions $D_0$ and $D_0'$ for $a>a_{\rm c}$ (we assume $D_0'<D_0$).
For $a=a_{\rm c}$, a single solution $D_0$ exists.
The critical amplitude $a_{\rm c}$ is determined together with the corresponding value $D_0$ of the diffusion coefficient by the two conditions
\be
f_0(D_0) = a_{\rm c} D_0^{\mu} \quad
f_0'(D_0) = \mu a_{\rm c} D_0^{\mu-1}.
\ee
Graphically, this correspond to the fact that for $a=a_{\rm c}$, the curves representing the functions $f_0(D)$ and $g(D)=a\,D^{\mu}$ intersect at a single point $D_0$ and have a common tangent at this point.

In addition, another solution can be found by assuming a scaling
$D=u|h|^{\nu}$, with now $\nu>1$. One finds to leading order for $h\to 0$,
\be
f(u|h|^{\nu},h) = u |h|^{\nu-1} + o(|h|^{\nu-1}).
\ee
Hence the solution of Eq.~(\ref{eq:self:consist:bne1}) is given for $h\to 0$ by
\be
\nu = \beta\,, \quad u = a^{\beta} = \alpha\,,
\ee
so that $D=D_h\equiv \alpha |h|^{\beta}$; in other words, $\Gamma=|h|$ as seen from Eq.~(\ref{eq:rel:D:Gamma}).
For $a<a_{\rm c}$, we thus have a single solution $D_h$ (with $D_h\to 0$), while for $a>a_{\rm c}$ we have three solutions $D_h<D_0'<D_0$. Physical intuition suggests that the intermediate value $D_0'$ may be unstable, while $D_h$ and $D_0$ may be stable, by analogy with the mean-field Ising phase transition for instance.
Yet, the stability of the three fixed points is difficult to assess analytically using the evolution equation (\ref{eq:evol:pxt}) supplemented by condition (\ref{eq:self:consist}) on the diffusion coefficient, as one would need to determine the time-dependent distribution $p(x,t)$.
As a simplified stability analysis which is expected to provide some
hints on the true stability properties, we propose to define instead a slow dynamics of the diffusion coefficient as follows:
\be \label{eq:evol:D}
\frac{dD}{dt} = -\zeta D + \zeta \alpha \Gamma(D)^{\beta} \equiv \zeta F(D)\,.
\ee
This dynamics of $D$ coupled to Eq.~(\ref{eq:evol:pxt}) shares the same stationary state as the original dynamics given by Eqs.~(\ref{eq:evol:pxt}) and (\ref{eq:self:consist}). Assuming the relaxation rate $\zeta$ to be small, one can use a quasi-stationary state approximation by plugging the slowly time-dependent diffusion coefficient $D(t)$ into the stationary solution $p_{\rm st}(x)$.

The solutions $D_0'$ and $D_0$ ($D_0'<D_0$) satisfy
$f_0(D_0')=g(D_0')$ with $f_0'(D_0')<g'(D_0')$,
and $f_0(D_0)=g(D_0)$ with $f_0'(D_0)>g'(D_0)$ (see Fig.~\ref{fig:scplot}, where the full line represents $f_0(D)$ and the dot-dashed line corresponds to $g(D)$ in the case of interest here).
The function $F(D)$ introduced in Eq.~(\ref{eq:evol:D}) can be rewritten as
\be
F(D) = D\left[ -1 + \left(\frac{g(D)}{f_0(D)}\right)^{\beta}\right]
\ee
so that $D_0$ and $D_0'$ are indeed fixed points of the dynamics given in Eq.~(\ref{eq:evol:D}). The stability of these fixed points is determined by the sign of $F'(D_0)$ and $F'(D_0')$.
For a fixed point $D^* \in \{D_0,D_0'\}$, one finds
\be
F'(D^*) = (\beta-1) \left[1-\frac{f_0'(D^*)}{g'(D^*)} \right].
\ee
It follows that $F'(D_0)<0$ and $F'(D_0')>0$: $D_0$ is a stable fixed point, and $D_0'$ is an unstable fixed point.
The instability of the fixed point $D_0'$, which satisfies $D_h < D_0' < D_0$, also implies from the one-dimensional character of the flow of $D$ that $D_h$ is a stable fixed point.

The average position $\la x \ra$ behaves very differently for the two fixed points. For the fixed point $D_0$, corresponding to a finite diffusion coefficient, the average position behaves again as in Eq.~(\ref{eq:xav:regular}).
In contrast, for the fixed point $D_h = \alpha |h|^{\beta}$ (with $\beta>1$), the diffusion coefficient goes to zero faster than $h$, and the assumption
$h/D \to 0$ used to derive Eq.~(\ref{eq:response:Dcst}) breaks down.
However, the situation is physically quite clear. Diffusion is very inefficient to counteract the effect of the external field $h$, while the resetting process is much faster than both diffusion and bias. It follows that the stationary
distribution $p_{\rm st}(x)$ becomes sharply peaked around $x={\rm sgn}(h)$,
and thus $\la x \ra \to {\rm sgn}(h)$ when $h\to 0$.

\section{Conclusion}
    \label{sec:conclusion}

    In this paper, we have shown that a population of random walks with stochastic resetting such that the diffusion coefficient is an increasing power law of the average resetting rate of the population exhibits a rich phenomenology, with phase transitions between a linear and a non-linear response to a small external field of the average position of the walk.
    When the diffusion coefficient is proportional to the average resetting rate ($\beta=1$), the proportionality coefficient $\alpha$ plays the role of a control parameter and the phase transition between linear and non-linear response occurs as function of $\alpha$, the non-linear regime corresponding to the low-$\alpha$ phase. This case had been previously investigated in the specific context of the deformation of soft amorphous materials (the HL model \cite{HL98,EPJE15}), where the random walk refers to the diffusion of the local mechanical stress. We have reformulated here the problem in the more general and abstract framework of diffusion with stochastic resetting.
    In addition, we have generalized the model to explore dependencies of the diffusion coefficient on the average resetting rate that were considered as unphysical in a soft matter context and thus not explored in this specific setting.
    For a sublinear dependence of the diffusion coefficient on the average resetting rate ($\beta<1$), diffusion is strong enough to induce a linear response whatever the proportionality coefficient. In contrast, for a superlinear dependence ($\beta>1$), diffusion may become so weak that the average position actually diverges in the zero external field limit. However, for a sufficiently large proportionality coefficient $\alpha$, a second stable solution emerges, with a finite diffusion coefficient leading to a linear response for $h\to 0$.

    We have focused here on the stationary state of the model. A rich phenomenology can also be found by looking at the time-dependent behaviour of the model, as investigated in \cite{Sollich17,Liu17} in a soft matter context with a linear dependence of the diffusion coefficient on the average resetting rate ($\beta=1$). Investigating the time-dependent behaviour of the model for general values of $\beta$ could be of interest for future work.

To conclude, this work describes a minimal way to introduce mean-field interactions in a population of random walks with stochastic resetting. We hope it may find applications beyond the previously studied soft matter problem, and that it may trigger further theoretical works on the more general problem of interacting random walks with resetting. Note that while the diffusion coefficient was assumed here to depend on the average resetting rate of the population, one may also assume that the external field depends on the resetting rate, as done for instance in the context of macroeconomic agent-based models \cite{Gualdi15}.

%\section*{Acknowledgments}

\bigskip

%%%%%%%%%%%%%%%%%%%%%%%%%%%%%%%%%%%%%%%%%%

\end{document}